# CABLE DYNAMICS APPLIED TO LONG-LENGTH SCALE MECHANICS OF DNA

S. Goyal, T. Lillian, N.C. Perkins and E. Meyhöfer
Department of Mechanical Engineering
University of Michigan
2350 Hayward, Ann Arbor MI-48109-2125 (U.S.A.)
ncp@umich.edu

*Abstract*

*This paper introduces the use of cable dynamics models as a means to explore the mechanics of DNA on long-length scales. It is on these length scales that DNA forms twisted and curved three-dimensional shapes known as supercoils and loops. These long-length scale DNA structures have a pronounced influence on the functions of this molecule within the cell including the packing of DNA in the cell nucleus, transcription, replication and gene repair. We provide a short background to the mechanics of DNA and suggest the logical connection to the mechanics of a low tension cable. A computational model is then summarized and example results are presented for DNA supercoiling and looping.*

## 1. BACKGROUND

Deoxyribonucleic acid (DNA) is a long chain biopolymer molecule that has been characterized [1] as "the most central substance in the workings of all life on Earth." Located within the nucleus of our cells, DNA contains the coded (genetic) information needed to synthesize all proteins and thus sustain life. Duplication, referred to as replication, and segregation of DNA are used to faithfully transfer this genetic information from one cellular generation to the next. These major biological functions of DNA follow not just from its chemical make up but also from its physical 'structure'. By structure, we refer to the often complex shape and state of stress of this long molecule and how they ultimately affect its biological functions. To get started, we need to first describe the basic chemistry and structure of DNA, the multiple length-scales involved, and the major biological functions that DNA performs. In doing so, we will also discuss why we believe it is promising to study the long-length scale mechanics of DNA by employing methods and models from the field of cable dynamics.

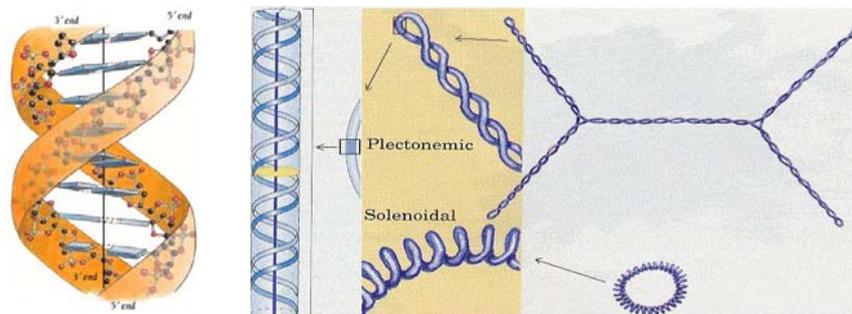

Figure 1: DNA shown on three length scales. Smallest scale (left) shows double-helix structure (sugar-phosphate chains and base-pairs). Intermediate scale (middle) shows how several double-helices form a continuous strand of DNA. Largest scale (right) shows how the strand ultimately curves and twists in forming supercoils (one interwound or plectonemic, and one solenoidal). Illustrations from Calladine and Drew [1]; Branden and Tooze [2]; Nelson and Cox [3].

Figure 1 illustrates a DNA molecule on three different length scales as reproduced from several sources [1-3]. The smallest length scale (far left) shows a segment of the familiar 'double-helix' which has a diameter of approximately 2 nanometers (nm). One complete helical turn is depicted here and this extends over a length of approximately 3 nm. The double helices, which wind like the supports of a spiral staircase, are composed of two polynucleotide chains which in turn are made up of four different nucleotides. Each nucleotide is made from a five-carbon sugar to which one or more phosphate groups and a nitrogen containing base are attached. There are four types of bases that include adenine (abbreviated A), guanine (abbreviated G), cytosine (abbreviated C) and thymine (abbreviated T). The four bases bond in only two unique, complementary pairs, namely A with T and C with G. The sugar-phosphate groups of the nucleotides are covalently linked into long chains (highlighted in orange) that form the backbone of DNA. Pairing of the two polynucleotide strands is achieved by hydrogen bonding between the nucleotide bases (highlighted in blue) that fill the small voids between the single DNA strands. It is this linear sequence of base-pairs that constitute the genetic code. Within the small voids between these chains lie the 'base-pairs' (highlighted in blue) that constitute the genetic code. There are four types of bases that include adenine (abbreviated A), guanine (abbreviated G), cytosine (abbreviated C) and thymine (abbreviated T). The four bases bond in only two unique pairs, namely A with T and C with G. This chemical structure and the rules for 'base-paring' follow from the seminal discoveries of Watson and Crick [4] and others. The base-pairs are hydrophobic and therefore must avoid contact with the surrounding aqueous environment within the cell. To this end, the double-helices effectively wrap around the base-pairs, thereby shielding them from the surrounding water molecules [1]. There are approximately 10.5 base-pairs in one helical turn for the common "B" form of DNA which also forms a right-handed helix as depicted in Fig. 1.

On an intermediate spatial scale (middle of Fig. 1), the double helix appears as a solid "strand" of DNA that might extend over tens to hundreds of helical turns (approximately tens to hundreds of nanometers). This is the approximate length scale of a 'gene' which is a portion of a DNA strand (i.e. a specific base-pair sequence) that controls a discrete hereditary characteristic. The base-pair sequence within a gene constitutes a chemical code for the production of a specific protein elsewhere within the cell. The major biological function of DNA is to store these chemical codes and to make them available for protein production through a process known as *transcription*. In addition, the same chemical codes are passed from one cell generation to the next through a process known as *replication*. Thus, transcription and replication are key biological processes essential for the functions of DNA. Transcription and replication are strongly influenced by the structure of the molecule on even longer length scales.

The human genome contains about 3.2 billion nucleotides organized into 24 different chromosomes. The total length of our DNA is about 1 m, which is about five orders of magnitude larger than a typical cell. These observations confirm that DNA is an exceedingly long (and flexible) molecule. The long-length scale structure of DNA is illustrated to the far right in Fig. 1. Here the long DNA strand may contain thousands to millions of base-pairs and resemble a highly curved and twisted filament with lengths ranging from micron to millimeter scales. The long-length curving/twisting of this strand is called *supercoiling* and two generic types of supercoils are illustrated to the far right of Fig.1. One type, referred to as an *interwound supercoil* (or plectoneme), leads to an interwoven structure where the strand wraps upon itself with many sites of apparent 'self-contact'. By contrast, a *solenoidal supercoil* possesses no self-contact and resembles a coiled spring or telephone cable. With the aid of proteins, DNA must supercoil for several key reasons. First, supercoiling provides an organized means to compact these very long molecules (by as much as $10^5$) enabling them to fit within the small confines of the cell nucleus. An unorganized compaction would hopelessly tangle the strand and render it useless as a medium for storing the coded information. Second, supercoiling may play an important roles in the biological processes of transcription and replication. For instance, the formation of simple loops of DNA on long-length scales is known to regulate the transcription of certain genes as we shall detail later in this paper.

It is on this largest length scale that DNA starts to resemble a (minute) cable. Consider for instance the striking similarities of DNA loops and supercoils to the loops and tangles (hockles) that form in low tension cables like those illustrated in Fig. 2. The models and methods used to understand how loops and tangles form in cables provide a natural means to explore the looping and supercoiling of DNA as described next.

## 2. RELATION TO CABLE/ROD MODELS

On its longest length scale, the proportions of a DNA molecule truly appear to be cable-like. Consider that [1] "The DNA from the longest individual human chromosome, if it were enlarged by a factor of $10^6$, so that it became the width of ordinary kite string, would extend for about 100 km." Such a long and slender molecule could indeed be modeled as a minute cable element provided one incorporates the specialized physical laws that are dominant at these length scales. Moreover, the curved and twisted structures that appear on long-length scales suggest the important roles played by the bending and torsion of a DNA strand. Cable models that capture bending and torsion employ rod theory. Indeed, the use of rod theory is reasonably well-established in the literature on DNA modeling as reviewed by Schlick [5] and Olson [6].

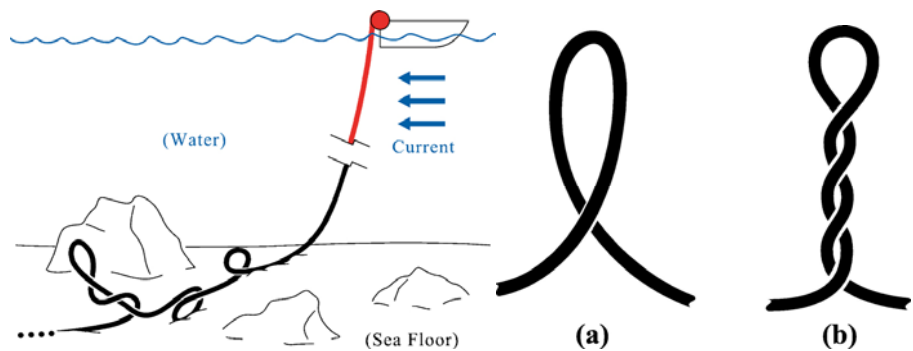

Figure 2: Marine cables under low tension can form *hockles* (or tangled loops) when subject to small (residual) torsion. Despite great differences in scales, hockle formation in cables is topologically equivalent to the formation of interwound supercoils in DNA.

While rod models may be naturally suited for describing the mechanics of DNA on long-length scales that extend over tens to millions of base pairs, they cannot describe the fine-scale structure of DNA at the base-pair level. Such fine-scale models of DNA can only be resolved through atom-by-atom descriptions of the DNA duplex (and the surrounding water molecules and any bound proteins/agents). However, the resulting molecular dynamics (MD) models rapidly grow to huge proportions and this limits their utility to very short (e.g., picosecond) time scales and to very short (e.g., nanometer) length scales. Thus, full molecular dynamics models cannot be used to simulate the long-length scale looping and supercoiling of DNA; see, for example, [7,8]. Other modeling techniques do exist (e.g., Langevin dynamics, Brownian dynamics, discrete link/chain models) that provide alternatives to MD simulation, see, for example studies reviewed in [5,6,9].

Numerous studies have employed rod theories to describe supercoiling of DNA under *equilibrium conditions* [10-24] starting with Benham [10, 11] who uses a hyperelastic, isotropic rod. The use of an isotropic (circular) rod to represent the structure of the double helix is specifically addressed by Maddocks and co-workers [12,13] who conclude that bending anisotropy at the base-pair scale quickly averages to an effective isotropic rod on long-length scales due to the high intrinsic twist (10.5 base-pairs/per helical turn) of the double-helix. Non-homogeneity (base pair sequence-dependent geometry and stiffness) in rod models is addressed in [14]. The studies [10-24] have contributed a fundamental understanding of the equilibrium states that describe supercoiled geometries (solenoidal and interwound), the stability of these states, and the physical parameters that control their bifurcations [15, 18-23]. Much of this understanding derives from the fact that, in the absence of body-forces, the governing equilibrium equations are integrable which greatly aids subsequent bifurcation analyses. Modeling the mechanics of interwound supercoils requires formulating "self-contact" in rod theory and this challenge has only recently been addressed [16,17, 24] in the context of closed loops (DNA "plasmids") [16,17] and long strands [24].

As noted above, the formation of supercoils in DNA is topologically equivalent to the formation of loops or *hockles* in marine cables as noted in [5, 21, 24, 25]; refer again to Fig. 2. Thus, it is appropriate to review the prior studies of cable looping and tangling (hockling) in this context. The earliest studies of cable hockling [26-28] employ equilibrium rod theory to evaluate the cable torque and tension required to initiate a "looping instability" and the converse "pop-out" instability which destabilizes the cable loop. Extensions that incorporate three-dimensional equilibrium forms, their local stability, and spatial complexity are provided in [25, 29,30]. A recent summary of the bifurcations responsible for looping and pop-out in twisted rods with clamped ends is presented in [24] together with compelling experimental results on (macro-scale) metal-alloy rods.

The studies cited above all employ *equilibrium* calculations to predict supercoiled states of DNA or hockled states of cables. An exception is the work of Klapper [9, 31] who formulates a dynamical extension to show how interwound equilibrium supercoils develop quasi-statically in one DNA plasmid (closed loop of DNA). Fundamental dynamical phenomena of supercoils are left unaddressed, including the existence of multiple supercoiled states and the possible nonlinear transitions between these states [1]. The need for dynamic treatments using rod theory is recognized, but not pursued, in [16, 24] which both note the limitations of the equilibrium rod theories they employ. By contrast, dynamical theories for cables are more prevalent and have recently been employed to study the nonlinear dynamic evolution of hockles. For instance, Gatti and Perkins [32] demonstrate the highly dynamic collapse of an initially straight cable under compression and the resulting nonlinear transitions to looped states. This approach is extended by Goyal, Perkins and Lee [33] who simulate hockles that develop from torsional buckling. Thus, recent studies of cable dynamics provide a natural avenue to explore the dynamical behavior of DNA looping and supercoiling [34].

## 3. A CABLE MODEL FOR LONG-LENGTH SCALE DNA MECHANICS

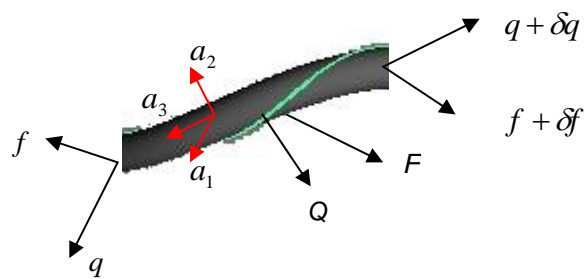

Figure 3: Free body diagram of an infinitesimal element of a DNA strand.

A general nonlinear dynamical model for a DNA strand is summarized below. This model develops from previous rod models for low-tension cables and includes the multi-physical effects needed at these length scales. These multi-physical effects include: 1) large deformations (rotations), 2) non-homogenous and non-isotropic behavior (e.g., base-pair "sequence-dependent" properties), 3) electrostatic 'self-contact' and interwinding, and 4) dissipation and thermal excitation from the aqueous solution. The implementation of these effects in a computational algorithm is also briefly described. A detailed derivation and discussion of this model can be found in [35].

Consider the element of a DNA strand shown as a free body in Fig. 3. Let the triad $\{a_i\}$ define a body-fixed reference frame fixed to the strand cross-section where $\hat{a}_3 = \hat{t}$ is the unit tangent vector to the centerline [36]. The quantities $Q$ and $B$ denote an external moment and force per unit length, respectively, while $q$ and $f$ denote the resultant internal moment and force, respectively, that act on the cross-section. Let the Lagrangian variable $s$ define the position of a material point on the strand centerline.

Four vectors are required to define the dynamic state of the cross-section and the internal stress resultants. These include the linear velocity $v$ of the centerline, the angular velocity $\omega$ of the cross-section, the curvature $\kappa$ of the centerline, and the internal force $f$. The kinematical quantities $\omega$ and $\kappa$ are smooth and are related by[1]

$$\frac{\partial \omega}{\partial s} + \kappa \times \omega = \frac{\partial \kappa}{\partial t}.$$

The centerline is inextensible to first approximation[2] which leads to the further requirement

$$\frac{\partial v}{\partial s} + \kappa \times v = \omega \times \hat{t}.$$

The balance law for linear momentum of the element is

$$\frac{\partial f}{\partial s} + \kappa \times f = m\left(\frac{\partial v}{\partial t} + \omega \times v\right) - F,$$

and that for angular momentum is

$$\frac{\partial q}{\partial s} + \kappa \times q = I\frac{\partial \omega}{\partial t} + \omega \times I\omega + f \times \hat{t} - Q.$$

Given requisite boundary and initial conditions, these four vector equations can now be used to evaluate the four unknowns $(v, \omega, \kappa, f)$ for the nonlinear dynamics of a strand subject to any prescribed external forces $F$ and moments $Q$ as in [33-35]. The quantities $F$ and $Q$ are, in general, nonlinear functions of the state vectors and they are used to capture the effects of element self-contact (e.g., interwinding) and the surrounding aqueous solution. The inertia properties of the element are described by the element mass/length $m$ and the inertia tensor/length $I$ for the strand cross-section about the triad $\{a_i\}$. A constitutive law for two-axis flexure and torsion must also be introduced. Prior studies of DNA mechanics [10-24] have employed linear material models based on results from single molecule experiments. Our formulation accommodates non-homogenous material properties that might capture, for instance, the base-pair "sequence-dependent" stiffness of strands and also any intrinsic curvature/twist. The constitutive law can also capture the chirality (i.e., a non-isotropic behavior) of the molecule that couples tension and torsion as detailed in [35].

---

[1] All the vector quantities are described in terms of components along the body-fixed unit axes $\{a_i\}$.
[2] This approximation can be relaxed and replaced by an appropriate constitutive relation for centerline extension. However, prior static analyses of DNA supercoils [12-25] suggest that extension plays a negligible role in the formation of supercoils in comparison to the dominant roles of flexure and torsion.

A significant challenge in predicting supercoiling is to account for the electrostatic interaction of remote segments of DNA as they approach and eventually interwind; refer to Fig. 1. We developed an efficient computational strategy to search for "self-contact" sites in [33]. A contact (repulsive) force is introduced between a pair of computational nodes only if two conditions are met: 1) the separation between the nodes lies within a specified tolerance, and 2) the nodes lie within each other's *conical aperture*. As discussed in [33], an aperture of angular width $\theta$ is constructed at each node as a pair of conical surfaces. These surfaces eliminate from consideration non-physical interactions between closely spaced grid points on the same segment thereby substantially reducing the numerical search for potential contact sites. Example interaction laws that can be employed include (attractive-repulsive) Lennard-Jones type [37], (screened repulsion) Debye-Huckle type [38], general inverse-power laws [31, 33], and idealized contact laws for two solids [16, 24].

DNA survives in an aqueous environment and any dynamic response is significantly damped by attached/surrounding water molecules. We capture this damping by employing a Stoke's regime drag model starting with published coefficients for the skin friction and the form drag for biomolecules [39]. A means to include thermal excitation is also suggested in [35].

The above theory is discretized using finite differencing methods that have been proven efficient for fluid-loaded cables [40]. In particular, we employ the Generalized-$\alpha$ method used in [33-35] to achieve a method that is unconditionally stable and $2^{nd}$ order accurate in both space and time. The resulting difference equations are implicit and their solution requires satisfaction of boundary conditions. An example shooting method algorithm for satisfying boundary conditions is described in [33-35].

## 4. EXAMPLE RESULTS

In this section, we review example results from our previous and ongoing studies of DNA supercoiling and looping. We begin with an early study of DNA supercoiling that illustrates several capabilities of our formulation including the ability to dynamically track sites of self-contact during the evolution of an interwound supercoil. Next, we provide an overview of recent computations of DNA loops and explore the energetics of this process for a DNA/protein complex found in the bacterium *E. coli*.

**Evolution of Interwound Supercoil**

As discussed in the Background, DNA often exists in supercoiled states for a variety of reasons including the need to pack this long molecule in an organized manner within the small confines of the cell nucleus. From a mechanics perspective, the DNA strand must curve and twist to a large degree in arriving at these supercoiled states. Among the many fascinating issues to explore are how supercoils might form, the energy required for their formation, and possible large-scale transitions from one supercoiled state to another. Here we will focus on one example that illustrates the evolution of a supercoil in an otherwise straight strand.

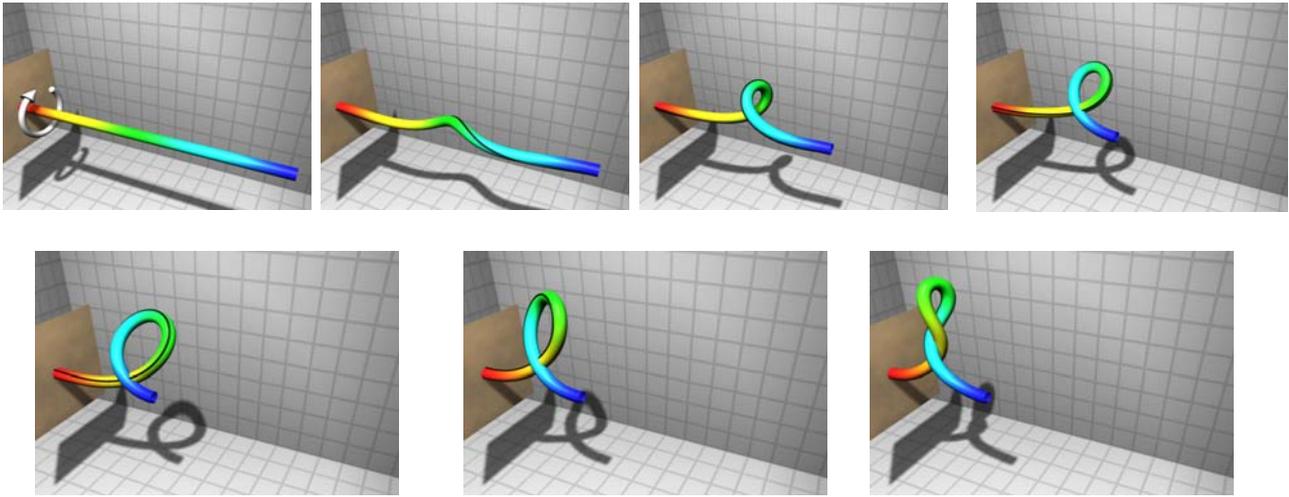

Figure 4: Results from Goyal, Perkins and Lee [33, 34] illustrating the evolution of an interwound supercoil (a plectoneme) under slowly increasing twist applied at the left end. The right end is restrained in rotation but is allowed to translate towards the left end.

Consider the sequence of numerical results illustrated in Fig. 4 for an idealized strand that is subjected to a slowly increasing torque. This strand is initially in a relaxed state that is also assumed to be perfectly straight. The torque is applied along the tangent to the left end and this end is otherwise constrained from all other rotation and translation (i.e., it is a fixed end that is only allowed to rotate about the tangent due to the applied end torque). The right end is fully constrained in rotation but it is allowed to translate towards the left end as the strand deforms. At the start of this process (first image top row), the applied torque produces a slow "winding up" of the strand without any lateral deformation. At a critical value of this torque, a bifurcation (torsional buckling instability) is reached and the strand deforms modestly into the approximate shape of a shallow helix (second image top row). The lateral deformation in this state is greatest near the center of the strand and, as the torque builds, so does the deformation in this zone as it ultimately produces a distinctive loop (third image top row). The plane of this loop rotates as this process continues and a secondary bifurcation occurs when this plane (actually the tangent at the midpoint) becomes orthogonal to the original axis of the strand (again third image top row). At this instant, the loop experiences a sudden dynamic collapse to a nearly planar form with one site of self-contact (last image top row). Increasing the torque further causes this loop to rotate about its axis of symmetry as it interwinds and multiple sites of self-contact are born as illustrated by the images in the second row.

There are several major challenges in resolving the solutions above in the context of a cable dynamics model. First, one must formulate a (numerical) method to track possible sites of self-contact. Second, one must introduce a local 'contact model' that captures the repulsive interaction of the two negatively charged portions of the strand in contact. Third, one must numerically stabilize the integration across the highly dynamic transitions from separation to contact. These challenges and the computational means to address them are discussed in further detail in [33, 34]. In addition, careful benchmarking of solutions is provided in [35]. Taken together, these results confirm the basic capabilities of the computational rod model in describing the dynamics of highly twisted and curved strands. We now move further to describe how this computational model can be used to explore a specific and fundamental biological mechanism known as protein-mediated DNA looping.

**Protein-Mediated DNA Looping**

In the Background, we described the essential function DNA plays in the production of proteins. Each protein is formed by a specific combination of amino acids and the 'code' for which amino acids and their order are specified by the base-pair sequence of an associated (specific) gene. The biological process known as transcription refers to the 'reading' of the base-pair code within a gene for the purpose of protein production [1]. The mechanisms that control transcription represent open and fundamental research issues in molecular biology.

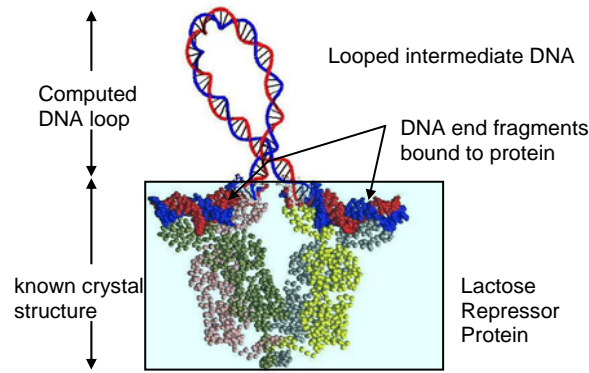

Figure 5: Protein-mediated DNA looping is a well-recognized mechanism for regulating the transcription of genes. The example system is a well-known gene from the bacterium *E. coli* that is responsible for the production of the enzyme lactose. When the "lactose-repressor" (Lac-R) protein binds to two operator sites on either side of this gene, a DNA loop is formed and transcription is repressed.

One such mechanism is the long-length scale looping of DNA as observed in the bacterium *E. coli*. Here transcription of a certain gene is repressed whenever a DNA loop is formed that contains that gene. The loop is formed by a protein that binds to two sites on either side of the gene. The forces produced by the relatively stiff protein cause the flexible DNA strand to form a loop as depicted in Fig. 5.

Crystallographic experiments of this DNA/protein complex can characterize the protein, and the small DNA fragments bound to the protein, but such techniques cannot reveal the structure of the intermediate loop of DNA. Our computational model can predict this looped structure by using the known location and orientation of the protein binding sites as boundary conditions for the rod model [41]. The predicted loop geometry, and the energy required to form this loop, are of considerable interest in understanding this fundamental gene regulation mechanism.

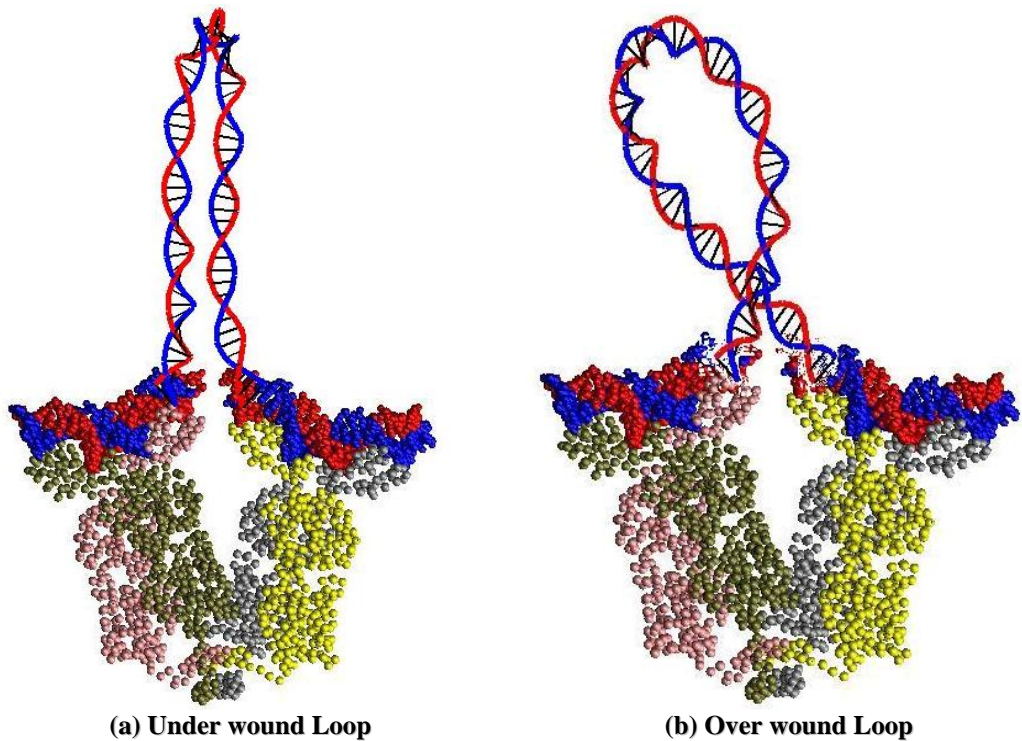

**(a) Under wound Loop**  **(b) Over wound Loop**

Figure 6. Predicted DNA loops for Lac-R/DNA complex. Two solutions for the loop are shown above with that in (a) being under wound and that in (b) being over wound. These predictions ignore any intrinsic curvature of the DNA strand as also assumed in [41].

Figure 6 illustrates example results obtained from the computational rod model as applied to this protein-DNA complex. The strand length is 75 base-pairs (approx. 25 nm) between the boundaries (protein binding sites). The elastic properties of the strand (bending and torsional stiffnesses) are approximated from experimental results performed on single molecule DNA as described in the literature; see, for example, [42-50]. The resulting nonlinear boundary-value problem is solved dynamically starting with an assumed straight strand of DNA in its relaxed state. Subsequently, the ends of the strand are translated and rotated into their final (known) position and orientation as defined by the experimentally-determined position and orientation of the protein binding sites as in [41]. Multiple final looped states are possible and these arise in the numerical computations by varying the sequence of end translations and rotations leading to the (same) final boundary conditions. The solution shown in Fig. 6(a) describes a loop that is 'under wound' in that it is slightly less twisted than the nominal (relaxed) DNA strand (i.e., the twist is slightly less than 10.5 base-pairs per helical turn). By contrast, the loop of Fig. 6(b) is 'over wound', that is slightly more twisted than the relaxed strand. This distinction is important biologically since less mechanical energy (work) is required to form the under wound loop in this case. The energy required for loop formation is an energetic cost that must be overcome by the protein/binding process in order to regulate transcription. In this case, the lowest energy (under wound) loop requires approximately 33 $K_B T$ for formation, where $K_B$ denotes the Boltzmann constant and T denotes (absolute) temperature. Overall, these predictions, which assume that the relaxed strand of DNA is straight, confirm prior predictions of loop geometry and energy as described in [41].

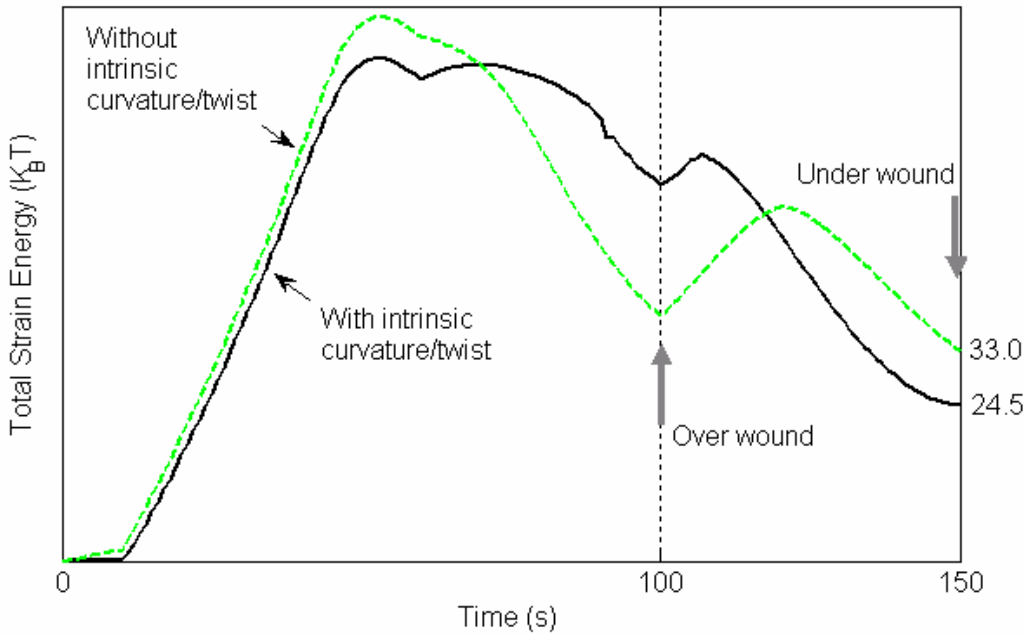

Figure 7: The energetics of loop formation for the DNA/Lac-R complex. The total strain energy is plotted as a function of time during the formation of both over wound and under wound loops. The green dashed curve shows the predicted results based on ignoring intrinsic curvature/twist while the solid black curve captures this influence. The lowest energy loops remain under wound however the looping energy is reduced by approximately 25% upon the inclusion of intrinsic curvature/twist.

Next, we demonstrate how these predictions change when one incorporates the actual (non-trivial) geometry of the relaxed DNA strand in the formulation. The relaxed DNA strand possesses intrinsic curvature and twist which are also dependent on the base-pair sequence. For instance, sub-domains that are rich in A-T base-pairs may introduce considerable intrinsic curvature known as 'A-tract bends'. Mechanically, additional bending in the same direction of the A-tract bends will be energetically more favorable (cost less) than bending opposite the A-tract bends. Beginning with the known base-pair sequence, one can systematically compute the intrinsic curvature and twist of the relaxed DNA strand by first employing the web tool [51] to determine the locations of all atoms in the strand. Next, one can fit a smooth (three-dimensional) curve through the center of each base-pair in arriving at the intrinsic curvature and twist of the relaxed strand. Incorporating this 'sequence-dependent' data in to the rod model renders the rod constitutive law non-homogenous but otherwise does not alter the computational strategy. The changes in the predicted results however are significant as discussed below.

Figure 7 illustrates the strain energy (two-axis bending plus torsion) computed during the simulation of loop formation. Starting from an unstressed state, the simulation proceeds through the sequence of boundary condition positions and orientations that result in the over wound loop at approximately 100 seconds and the under wound loop at approximately 150 seconds. The results that capture intrinsic curvature/twist (shown in solid black) differ significantly from those that ignore this effect. While both predictions lead to the same conclusion that the under wound state is more energetically favorable, this energy is reduced by approximately 25% upon the inclusion of sequence-dependent intrinsic curvature/twist.

## 5. SUMMARY

This paper describes a new application of models developed for cables to describe the mechanics of strands of DNA of nanometer diameter and micron to millimeter length. Despite the large differences in length scales between a cable and a DNA molecule, their mechanics bear some striking similarities. In particular, the looping and tangling of low tension cables are topologically equivalent to the looping and supercoiling of DNA.

We review a computational rod model that has been developed for describing the looping and supercoiling of DNA on long-lengths scales. By 'long-length', we refer to lengths larger than a helical turn (3 nm) and as long as the millimeter scale. At such length scales, the strand resembles a long elastic rod (or cable) that may also become highly curved and twisted. The multi-physical effects at these length scales that can be captured in the rod model include: 1) large deformations (rotations), 2) non-homogenous and non-isotropic behavior (e.g., base-pair "sequence-dependent" properties), 3) electrostatic 'self-contact' and interwinding, and 4) dissipation and thermal excitation from the aqueous environment.

Example results are described from our previous and ongoing studies of DNA mechanics. We review results from an early study of supercoiling where we numerically explore how an interwound supercoil may evolve and the bifurcations that it experiences through a build up of twist. We then turn to recent results on DNA looping where we investigate the energetics of a specific DNA-protein complex found in the bacterium *E. coli*. Our computations suggest the critical role played by sequence-dependent intrinsic curvature as a means to lower the energetic cost of loop formation in this fundamental gene regulation mechanism.

**ACKNOWLEDGMENTS**

The authors gratefully acknowledge the research support provided by the National Science Foundation and by the Lawrence Livermore National Laboratories for our studies. We also acknowledge the fruitful discussions with our collaborators including Dr. C. L. Lee (Lawrence Livermore National Laboratories) and Professor C. Meiners and S. Blumberg (Biophysics, University of Michigan) and Professor Ioan Andricioaei (Biochemistry, University of Michigan).